\documentclass[letterpaper,11pt,twoside,onecolumn,final]{JIDPS}       
\usepackage{titlesec}								
\titlelabel{\thetitle.\quad}							
\usepackage[natbibapa]{apacite}						
\usepackage[margin=1in]{geometry}   				
\usepackage{mathptmx}				 				
\usepackage{amsmath}					 			
\usepackage{amssymb}
\usepackage{multirow} 					 			
\usepackage{epstopdf}					 			
\usepackage[dvips]{graphicx}             	 			
\usepackage{subfigure}                     					
\graphicspath{{Figures/}}                    				
\usepackage{algorithmicx, algpseudocode}			
\usepackage{enumerate}
\usepackage{indentfirst}                     				
\usepackage{setspace}                       				
\usepackage{algorithm}						   		
\usepackage{lineno,xcolor}							
\usepackage{fancyhdr}                       				
\usepackage[font={footnotesize,bf,sf}, labelfont={sf,bf}, justification=centering, labelsep=period, margin=5mm]{caption}	 
\usepackage{times}								
\usepackage{everypage}                      			
\usepackage{ifthen}                             			
\usepackage[bookmarksopen,bookmarksnumbered,colorlinks,urlcolor=blue,linkcolor=blue,anchorcolor=blue,citecolor=blue]{hyperref} 	
\newcommand*{\doi}[1]{\href{http://dx.doi.org/\detokenize{#1}}{ \detokenize{#1}}}	

\AddEverypageHook{
  \ifthenelse{\value{page}=1}
    {\setlength{\headheight}{65pt} \setlength{\headsep}{-40pt} } 
    {\setlength{\headheight}{13pt} \setlength{\headsep}{25pt} }} 

\begin{document}

\vspace*{24mm}

\begin{spacing}{2}

\noindent{\LARGE \bfseries
Time Reversed Delay Differential Equation Based Modeling Of Journal Influence In An Emerging Area}

\end{spacing}
\vspace*{4mm}

Poulami Sarkar$^1$, 
Snehanshu Saha$^1$\footnote{Corresponding author. Email: \href{mailto: snehanshusaha@pes.edu }{snehanshusaha@pes.edu}} , Archana Mathur$^2$,  Rahul Aedula$^1$, Saibal Kar$^3$, Surbhi Agrawal$^1$ and Kakoli Bora$^4$

\vspace*{2mm}

\noindent{\emph{$^1$PES Institute of Technology, Bangalore South Campus, Bangalore, India  \\
$^2$Indian Statistical Institute, 8th Mile, Mysore Road, Bangalore, India \\ $^3$Centre for Studies in Social Sciences, Calcutta, India
}}

\vspace*{10mm}

\noindent\textbf{Abstract} {
A recent independent study resulted in a ranking system which ranked Astronomy and Computing (ASCOM) much
higher than most of the older journals highlighting its niche prominence. We investigate the
notable ascendancy in reputation of ASCOM by proposing a novel differential equation based modeling. The
modeling is a consequence of knowledge discovery from big data-centric methods, namely L1-SVD. The inadequacy of
the ranking method in explaining the reason behind the growth in reputation of ASCOM is reasonable to understand
given that the study is post-facto. Thus, we propose a growth model by accounting for the behavior of parameters
that contributes to the growth of a field. It is worthwhile to spend some time in analyzing the cause and control
variables behind rapid rise in reputation of a journal in a niche area. We intend to identify and probe the parameters
responsible for its growing influence. Delay differential equations are used to model the change of influence on a
journal’s status by exploiting the effects of historical data. The manuscripts justifies the use of implicit control variables and models those accordingly that demonstrates certain behavior in the journal influence.
}

\vspace*{6mm}

\vspace*{4mm}


\section{Introduction}
\label{Sec.:One}

It is well-known that ranking of journals, whether in science, technology, engineering or in social sciences, such as in economics, is a contentious issue. For many subjects, there is no correct ranking, but a universe of rankings, each a result of subjective criteria included by its creators. In this regard, the following studies are instructive: (\citet{Kristie}, \citet{Jangid2014}). With the creators' choices and rules laid out explicitly, the users of such ranking still need to use own judgments and institutional requirements to choose ranks appropriately. The subjective element in journal rankings not only complicates matters about what is correct, if any, but also about outcomes that depend crucially on adoption and analysis of rankings. For science and related subjects, SCOPUS and SCIMAGO hold some of the best journal ranking systems to this day, using their Cite Score and SJR indicators respectively, to rank journals.(\citet{Kianifar2014}) However, owing to the manner in which both these indicators are considered, it is often the case that the received ranking might not always display the true quality and outreach of a specific scientific journal. Obviously, this could be true for a large number of subjects across length and breadth of contemporary research therefore recourse to a scientifically more acceptable method should always be of interest and often beneficial for a large set of users. To demonstrate this, we therefore considered the case of the Journal entitled, \textit{Astronomy and Computing}, within the context of SCOPUS Journals in the relevant domain of AstroInformatics (\citet{Bora2016}) , in particular and Astronomy and Astrophysics, in general.
\par  The primary focus of this case study is to determine the standing of Journal \textit{Astronomy and Computing} with respect to other journals which were established prior to it. \textbf{ More importantly, the reasons for such standing need to be investigated which is a more complex and qualitative study.} The algorithm also tests the validity of the ranking and suggests an alternative rank that used a more holistic approach towards the features. While this paper focuses on a specific journal, it is easy to see that the purpose of this construct is broad-based and deep-seated at the same time, such that the applications of the algorithms can be adopted by numerous other subjects grappling with the same problem.    
\begin{table*}[h]
\small{
\caption{Case Study: Astronomy and Computing, SJR (\citet{DBLP_sjr}) and L1-SVD ranks (\citet{l1SVD})}}
\begin{center}
\noindent \begin{tabular}{|c||c||c||c|}
 
\hline
\textbf{Journal Name} & \textbf{L1 Scheme Rank}& \textbf{SJR based Rank} & \textbf{Year}  \\ \hline
Astronomy and Computing & 39 & 31 & 2013  \\ \hline
Astronomy and Astrophysics Review
 & 40 & 5 & 1999\\ \hline
Radiophysics and Quantum Electronics &41 & 51 & 1969   \\ \hline
Solar System Research  & 42 & 48 & 1999\\ \hline
Living Reviews in Solar Physics  & 43 & 3 & 2005\\ \hline
Astrophysical Bulletin & 44 & 45 & 2010\\ \hline
Journal of Astrophysics and Astronomy  & 45 & 55 & 1999\\ \hline
Revista Mexicana de Astronomia y Astrofisica  & 46 & 23 & 1999\\ \hline
Acta Astronomica & 47  & 20 & 1999\\ \hline
Journal of the Korean Astronomical Society  & 48 & 32& 2009\\ \hline
Cosmic Research  & 49 & 58& 1968\\ \hline
Geophysical and Astrophysical Fluid Dynamics & 50  & 46 &1999\\ \hline
New Astronomy Reviews  & 51 & 12 & 1999\\ \hline
Kinematics and Physics of Celestial Bodies & 52 & 65 & 2009 \\ \hline
Astronomy and Geophysics  & 53 & 67 & 1996\\ \hline
Chinese Astronomy and Astrophysics & 54  & 72 & 1981\\ \hline
\end{tabular}
\end{center}
\end{table*}

\section{Motivation: The ranking scheme} 
We implemented $l_1$-norm SVD scheme using the publicly available SCOPUS dataset to rank all its corresponding journals, and simultaneously determine the potency of the algorithm. The outcome of the ranking scheme posed interesting and compelling questions which led us to model the growing influence of the particular journal. We discuss the detailed method in Appendix A, for the simple reason that the focus of the manuscript is not on the ranking methods, rather on	 the model formulation and interpretation explaining such rank.  SCOPUS contains approximately 46,000 Journals listed in different domains. Discarding few redundancies, SCOPUS effectively covers a large range of metrics and provides adequate resources for verification. For this demonstration, we have considered 7 different metrics from SCOPUS to be used as features in our algorithm. These features include \textit{Citation Count}, \textit{Scholarly Output}, \textit{SNIP}, \textit{SJR}, \textit{Cite Score, Percentile and Percent Cited}. \par
Indeed, to cross-verify the results of the algorithm these were compared to SJR based ranking of SCIMAGO for suitable articulation of the discrepancies. It seems that the $l_1$-norm SVD scheme works quite successfully (\citet{l1SVD}) in rating the journals and approaches the data in a more comprehensive way. The result is a ranking system which ranks \textit{Astronomy and Computing} much higher than most of the older journals and at the same time highlights the niche prominence of the particular journal. Similarly, this method also highlights the rise of other journals which were underrepresented due to the usage of the SCOPUS and SCIMAGO indicators only. This method, therefore, has been largely successful in rectifying the rank of such journals. Importantly, the $l_1$-norm SVD scheme can be extrapolated to other data as well. It can 	be used to study the impact of individual articles, for example. Utilizing similar features such as Total Citation, Self Citation, and NLIQ (\citet{Ginde2016}), the algorithm can be used to rank articles within a journal with great accuracy along with a holistic coverage. To re-apprise the scope of this research, it is important to remember that the common practice (\citet{Kristie})has been to control for the size of the journal (measures like pages, number of articles, even characters), age of article, age of citation, reference intensity, exclusion of self-citations, etc. 
\par In order to be precise, the ranking scheme raises some important questions which can be reasonably challenging. Standard scientometric features used to study influence/reputation of journals are not adequate for explaining the ascendancy of ASCOM in influence. The importance of investigating intrinsic dynamics is rarely stressed upon in scientometric literature (\citet{articleFei}). Usually, the analysis is static, based on citations and other factors. The authors intend to bring out the missing dynamics via the DDE based model. The following set of questions are addressed in this study. What are the non-quantitative factors (could be qualitative and difficult to quantify) explaining the rapid growth of this journal? What is the direction of causation, and how do we frame it? Does the big data landscape help? Can we formulate a model that reasonably accounts for such surge in influence? Are there features/factors, not statistically significant but play crucial roles as implicit control variables toward the phenomena?
The proposed model (Section 4 onward) addresses these questions. 
\subsection{Knowledge Discovery and the Evolution of ASCOM: Key Motivation for the model}
Albeit, Astronomy and Computing (ASCOM) has been in publication for five years only, its reputation has grown quickly as can be observed from the ranking system proposed here. This is despite the fact that ASCOM is severely handicapped in size. There is no journal focused on the interface of astronomy and computing in the same way as ASCOM. It can be observed from Table 1 that, ASCOM, in comparison with the other journals listed, is significantly younger! Unless the number of volumes and issues published are significant, a journal is unlikely to create the equivalent impact of an established journal. This is a notable handicap for any new journal, ASCOM being no exception. We define this as "size handicap". 
\par Despite the "size handicap" explained above, ASCOM is ranked 39 according to our method, slightly lower than its 31 rank in SCOPUS. This is due to the fact that we have not used "citations from more prestigious journals" as a feature (this data are not readily available). Nonetheless, it is ranked higher than many of its peers which have been in publication for over 20 years. This is also due to the fact that ASCOM is "one of its kind" and uniquely positioned in the scientific space steered by appropriate editorial support. However SUBJECTIVE the statement may sound, it seems that interdisciplinary, diversity in background of the Editors and authors  and novelty in theme have been instrumental in placing journals uniquely (\citet{NAKAWATASE2017}. \citet{Rodrguez2016} \citet{Jacob2012} \citet{Amin}). Such qualitative feature, regrettably is not visible from the big data landscape alone. This is another significant driving factor behind framing and interpreting a novel model that explains trends arising from investigating the big data landscape.
\begin{figure}[h]
 \includegraphics[width=8cm]{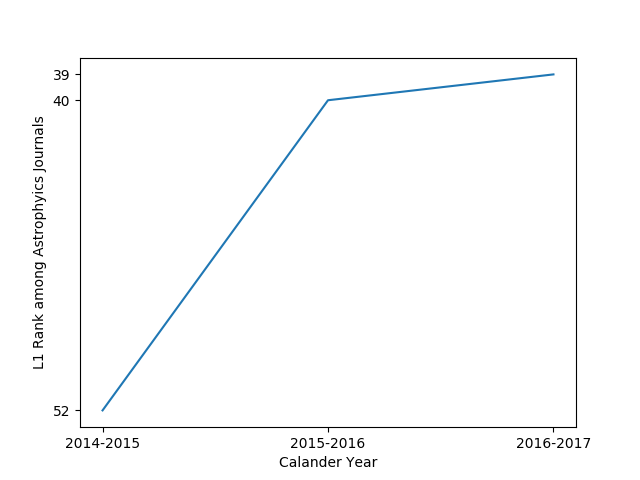}
 \centering \caption{$l_1$ Rank Progression of ASCOM based on SCOPUS data computed by the proposed method. Since, the steady ascendancy in the journal's rank is unmistakable, it will be interesting to investigate the behavior of the journal rank in the long run once enough data is gathered. Please see Appendix A for details.}
\centering
\label{fig:rank}
\end{figure}
\par There is another interesting observation to take note of. By ignoring the "size does matter" paradigm, the ranks of some journals (many years in publication with several volumes and issues) suffered according to our method. A few examples include Living Reviews in Solar Physics, ranked 43 according to our scheme while it is ranked 3 in SCOPUS; and Astronomy and Astrophysics Review, ranked 40 in our scheme while it is ranked 5 according to SCOPUS. This reversal of positions should be considered as important findings, because existing methods do not offer appropriate weights to journals that are new, despite catering to a niche and important area of research. In other words, the results indicate that years in publication may sometimes dominate over other indicators of quality and may not capture the growth of journals in "short time windows". Our study also reveals that ASCOM is indeed a quality journal as far as  early promise is concerned.
\\ Scientometrics deals with analyzing and quantifying works in science, technology, and innovation. It is a study that focuses on quality rather than quantity. The journals are evaluated against several metrics such as the impact of the journals, scientific citation, SJR, SNIP indicators as well as the indicators used in policy and management contexts. The practice of using journal metrics for evaluation involves handling a large volume of data to derive useful patterns and conclusions (\citet{bigdata}). These metrics play an important role in the measurement and evaluation of research performance. Due to the fact that most metrics are easily susceptible to manipulation and misuse, it becomes essential to judge and evaluate a journal by using a single metric or a reduced set of significant metrics. We proposed \textit{l$_1$}-norm Singular Value Decomposition(\textit{l$_1$}-SVD) (\citet{l1SVD}) to efficiently solve this problem. The code of the proposed method is available at \citet{git}.
\subsection{The Big Data Landscape}
The appeal of modern-day computing is its flexibility to handle volumes of data through an aspect of coordination and integration. Advancements in Big Data frameworks,(\cite{Apache}) and technologies has allowed us to break the barriers of memory constraints for computing and implement a more scalable approach to employ methods and algorithms.  The aforementioned journal ranking scheme is one such algorithm which thrives under the improvements made to scalability in Big Data. With optimized additions such as Apache Spark to the distributed computing family, the enactment of $l_1$ Regularization and Singular Value Decomposition has reached an all new height. Implementing the SVD algorithm with the help of Spark can not only improve spatial efficiency but temporal as well. The $l_1$-norm SVD scheme utilizes the SVD and regularization implementation of \textit{ARPACK} and \textit{LAPACK} libraries along with a cluster setup to enhance the speed of execution by a magnitude of at least three times depending on the configuration. Collecting data is also a very important aspect of Big Data topography. The necessity of a cluster based system is rendered useless without the requisite data to substantiate it. Scientometric data usually deals with properties of the journals such as Total Citation, Self-Citation etc. This data could be collected using Web Scraping methodologies but also can be found by most journal ranking organizations, available for open source use; SCOPUS and SCIMAGO. For the $l_1$-norm SVD scheme, we used SCOPUS as it had an eclectic set of features which were deemed appropriate to showcase the effectiveness of the algorithm. The inclusion of the two important factors such as Cite Score and SJR indicators gave a better enhancement over just considering one over the other. For more information about the data and code used to develop this algorithm (please refer to \cite{Github}, \href{https://github.com/rahul-aedula95/L1_Norm}{\underline{Github}} repository of the project).
\par The landscape and the framework, although attractive are insufficient to explain the rapid rate of growth of ASCOM. The remainder of the paper is organized as follows. We begin by presenting the motivation for Delay Differential Equation (DDE) based modeling by outlining key strengths of such modeling concept. Next, we consider time reversed DDE to model the growth by including historical effects, a fundamental contribution in section 4. Section 5 contains solutions, analytically and computationally investigated and interpreted in light of the big data landscape. Section 6 considers further modifications in the model by adding Editorial reputation and Publisher Goodwill value. We discuss the implications of these additional factors and the fundamental assumptions in Discussion \& Conclusion Sections, 7 and 8.

\section{Scope of our study and Motivation for modeling via DDE}

The manuscript strives to achieve two fundamental objectives:
\begin{itemize}
\item We establish and quantify current journal influence as a function of its past influence. If the past influence 
is positive (good inheritance), the present journal influence benefits immensely from it (Please see sections 4, 5 and fig.s 2, 3 and 4).
\item The manuscript proposes a doctrine of " self-serving incentivization" by exploiting implicit control variables  (publisher goodwill value and editorial reputation-the celebrity effect). The so-called " incentivized model" is 
proposed to propagate a positive "start-up boost" to the journal influence. Thereby, these control variables and the modifications form the second and more advanced, complex layer in modeling journal influence (Please see 
sections 6, 7 and fig.s 5-10) and help quantify the theory of " celebrity effect".
\end{itemize}

The factors mentioned above and the resulting model explained in the subsequent sections also account for the remarkable growth in influence and ASCOM discussed in sections 1 and 2. We achieve this by the DDE based model presented below.\\
\par DDE is a well known concept for over two centuries, which has found application in various problems in the fields of dynamical modeling of biomedical systems, biochemical reactions as well as in the newer models of interpersonal/romantic relationships!! DDEs also find useful applications like dynamic population growth, economic growth and spread of diseases like HIV, cancer, etc. Delay Differential Equations belong to the class of Partial Differential Equations. These are used by the scientific community for modeling dynamic systems for many of the obvious advantages. These equations describe the rate of change of a function, at time 't' as a function of earlier times. A DDE in its general form can be given by:
\begin{equation}
p'(t) = f(p(t),p(t-\tau)); p(0)=  p_0
\end{equation}
considering a constant delay of $\tau$.
\noindent
Some of the advantages of DDEs are:
\begin{itemize}
\item DDE take care of the "hereditary effects" during modeling a system. This implies if the influence of a journal is positive in the past and/or intrinsic factors have been responsible for surge in reputation, such features are naturally modeled in DDEs.
\item In system modeling, it is desirable that the model is closer to the real process (in our case, influence diffusion and percolation) and it has been observed that DDEs offer a better model than others.
\item DDEs are seen to provide a better control over the system since historical data is directly modeled in to the system (using time-reversed structure). This is particularly desirable.
\item In case of a DDE, the initial point $p_0$ defined over the interval [-$\tau$, 0] , is a function and not just a point. The solution $p(t)$ is also a function in the same interval. Hence, the solution becomes infinite dimensional, unlike an ODE. Moreover, in a dynamical system, DDE takes care of rate of growth,  which is a robust form of looking at the real world problem than just reading from hereditary events and inferring from them.
\end{itemize}
\section{TIME REVERSED DDE: Our Contribution}
Let $p'(t)$ ~ rate of change of influence over time, $p(t)$ ~ influence @ time $t$, and $p(-t)$ ~ influence @ time $t = -t$ ($t=1, p'(1) = ap(1) + bp(-1)$  or $p'(2) = ap(1) + bp(-2)$ and so on).
The Time Reversed equation can now be written as 
\begin{equation}
p'(t) = ap(t) + bp(-t)
\end{equation}
which implies the rate of change of influence is represented as a combination of present and past influence.
Let us consider a simple growth model given as 
\begin{align*}
ap'(t) &= b + cp(t) \\
p(0) &= c 
\end{align*}
where $p(0)$ is not the Initial condition but is the value at the instant of time under the interval of consideration. We represent this linear growth in the form of time reversed structures as follows:\\
\begin{align*}
\Longrightarrow p'(t) &= \frac{b}{a} + \frac{c}{a} p(t)\\
 &= \frac{b}{a} + \frac{c}{2a}p(t) + \frac{c}{2a}p(t) \\
&= \frac{b}{a} + \frac{c}{2a}p(t) + \frac{c}{2a}p(-t) 
\end{align*}
We assume symmetric influence function; there are two possibilities, symmetric and non-symmetric influence. Differentiating w.r.t $t$,
\begin{align*}
p''(t) &= \frac{c}{2a}p'(t) - \frac{c}{2a} p'(-t)\\
 &= \frac{c}{2a} \Big( \frac{b}{a} + \frac{c}{2a}p(t) + \frac{c}{2a}p(-t) \Big) - \frac{c}{2a}\Big(\frac{b}{a} + \frac{c}{2a}p(-t) + \frac{c}{2a}p(t)\Big) \\
&= 0
\end{align*}
Note: $p(t)$ may exhibit linear growth under the assumption that there is a certain repeatability in the journal influence. 
\subsection{The model under non-symmetric influence:}
Let us not consider the symmetric influence function since it is too strong an assumption to begin with (fluctuations are absent, unidirectional slope, elements of uncertainty almost absent). Let us consider the same model given as 
\begin{equation}
ap'(t) = b + cp(t);
p(0) = c 
\end{equation}
without the assumption of symmetric influence (p(t)=p(-t)). Here also, $p(0)$ is not the Initial condition but is the value at the instant of time under the interval of consideration.
Reorganizing equation 3, 
\begin{equation}
p'(t) + (-\frac{c}{a})p(t)=\frac{b}{a}
\end{equation}
Assuming $(-\frac{c}{a})$ and $(\frac{b}{a})$ are continuous functions, we fix $(-\frac{c}{a}) = r(t)$ and $(\frac{b}{a})=s(t)$. Putting this in the equation, we obtain
\begin{equation}
p'(t)+r(t)p(t)=s(t)
\end{equation}
Let $\mu(t)$ be an integrating factor(\citet{diff_equ}). Multiply both sides of equation with $\mu(t)$ and integrating, we arrive at the following form:
$$\mu(t)=Ke^{\int r(t)dt}$$
and eventually the expression for journal influence is written as 
\begin{equation}
p(t)=\frac{\int e^{\int r(t)dt}s(t)dt + \frac{c}{K}}{e^{\int r(t)dt}}
\end{equation}
Under the assumption of non-symmetric influence (more realistic), the influence seems exponential growth or decay depending on the coefficients but not a combination of both in a single expression. We shall see a different picture in the next section when we encounter non-linear growth in influence for a slightly more complicated, time reversed model.\\
\textbf{Remark:} Please note the above model does not contain "history" functions. Hence the solution does not display a convex combination of exponential functions, which can be easily interpreted in light of historical data. This is in contrast to the simple case (we assume a symmetric influence) where we can safely conclude that if either the historic influence or the current influence of the journal is high then the journal is most likely going to experience further rise in influence in the near future.
\subsection{Modeling Non-linear growth using symmetric influence effects}
Let us consider eq.(1) with the condition $p(0) = c$
by mapping these to the following DDE: 
\begin{align*}
y'(t) &= a_1(t)y(t) + a_2(t)y(t-d), t>=0\\
y(t) &= p(t), t \in [-d,0]
\end{align*}
Consider $d = -2t; a = a_2(t), b = a_1(t); y(t) \equiv p(t)   \forall t \in [-d, d]$. Our proposed model is a special case of DDE and it will be shown later that eq.$(1)$ has at least one solution, which may not be necessarily unique.\\
\textbf{Solution Methodology:} Let us consider the time reversed model eq.$(2)$:
\begin{align*}
p'(t) &= ap(-t) + bp(t)\\
p(0) &= k\\
p'(0) &= (a+b)k
\end{align*}
By Symmetry, we have 
\begin{equation}
p'(-t) = -ap(t) -bp(-t)
\end{equation}
Differentiating Eq.(2) wrt t, we get,
\begin{align*}
p''(t) &= ap'(-t) + bp'(t)\\
 &= a\big(-ap(t)-bp(-t)\big) + b\big(ap(-t) + bp(t)\big)\\
 &= -a^2p(t)-abp(-t)+bap(-t)+b^2p(t)\\
 &= (b^2-a^2)p(t) 
\end{align*}
where   $r = \sqrt[]{b^2-a^2}$

Again, by symmetry,
\begin{equation}
p''(-t) = r^2p(-t)
\end{equation}
Solution is of the form,
\begin{equation}
p(t) = A\exp(rt) + B\exp(-rt)
\end{equation}
It is evident that $p(t)$ is an exponential function.
Using initial conditions, solving for A \& B in terms of a \& b we get,
\begin{equation}
p(t) = \frac{c}{2r}\Big(r+a+b\Big)\exp(rt) + \frac{c}{2r}\Big(r-a-b\Big)\exp(-rt)
\end{equation}
Depending on the coefficient values, either positive or negative exponents will dominate.
The two possible solutions depend on the value of \textit{r}. 
\begin{itemize}
\item When $r > 0 (i.e., b > a)$, we can expect an exponential real solution
\begin{figure}[h]
 \includegraphics[width=9cm]{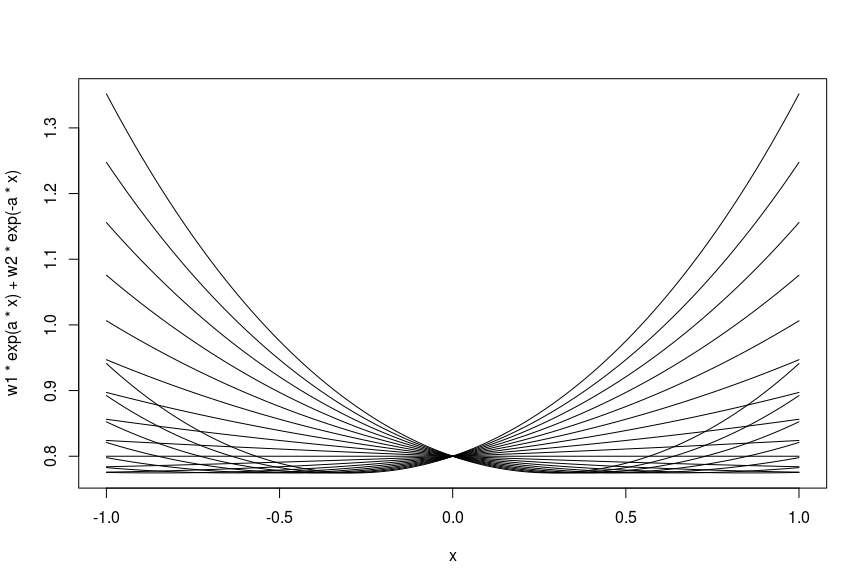}
 \centering \caption{Plot of eq. (9) where $p(t)$ is represented by the Y-axis and t is represented by the X-axis. The present influence,$p(t)$ is controlled by past influence, $p(-t)$}
\centering
\label{fig:rank}
\end{figure}
\item When $r < 0 (i.e., b < a)$, there will be oscillatory solutions, due to r being imaginary. Again, these solutions are deemed infeasible due to lack of fixed periodicity.
\begin{figure}[h]
 \includegraphics[width=8cm]{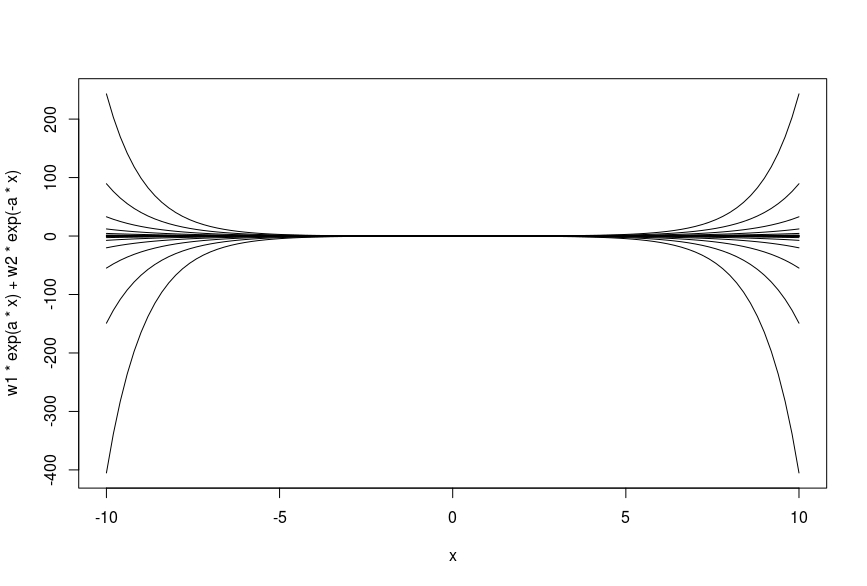}
 \centering \caption{Plot of eq. (9) where $p(t)$ is represented by the Y-axis and $t$ by the X-axis.Imaginary solutions are obtained when $a$ and $b$ are varied such that $w1 < 0$. This is infeasible as the model explains real solutions for obvious reasons.
}
\centering
\label{fig:rank}
\end{figure}
\end{itemize}
$t = 0$ is considered to be in the middle of a short time frame, at which, we are measuring the influence. Hence, this is not considered as initial value problem and hence we are not guaranteed of a unique solution. $p(-t)$ is the mirror image of $p(t)$ and it will result in a sharp spike in influence provided its value is high. This is typically observed in a short time window and averages out in the longer time span.
We see that depending on the values of the parameters and b either the historical or the current data dominates . The curve shows that in first few years the influence is largely dominated by the past reputation of the editors , represented by the historical part of the DDE. After a certain point ( we have assumed this point to be at the center of time series data), other parameters such as the current journal citations and the current reputation of the editors begin to reflect on  the influence.

\section{Model Fitting:}
Let us recall Eq.$(1)$:
\begin{align*}
p'(t) &= ap(t) + bp(-t)\\
p(0) &= c\\
p'(0) &= (a+b)c
\end{align*}
We also know, by approximation that,
\begin{align*}
p'(t) \approx \frac{p(t+h)-p(t)}{h}\\
\approx \frac{p(t)-p(t-h)}{h}
\end{align*}
where $h$ is the step size.
Let us consider the spread at discrete time intervals corresponding to one to five years (obtained from the dataset), indicated as $p'(1), p'(2), p'(3), p'(4) and p'(5) $ respectively.
Here, we can write $p'(1) = ap(-1) + bp(1) $. Also,
\begin{align*}
\Longrightarrow \frac{p(1)-p(0)}{1} &= ap(-1) + bp(1)\\
&= a[A\exp(-r) + B\exp(r)] + b[A\exp(r)+B\exp(-r)]\\
&= (aA+bB)\exp(-r) + (aB+bA)\exp(r)
\end{align*}
The value on LHS is obtained from the dataset. Similarly, we can compute $p'(\frac{3}{4}), p'(\frac{1}{2}),p'(\frac{1}{4}),$ etc., can be obtained from the dataset, where the fractions represent the quarters in a year. We are now required to estimate the coefficients a, b, A \& B. This is an overestimation problem with number of equations exceeding number of unknowns. We can solve this by method of Least Squares and hence use the solution to predict future influence in rate of journal influence spread.
\subsection{Least Square Method to fit the data:}
From eq. (1), we obtain
\begin{align*}
p'(t) &= ap(t) + bp(-t)\\
\end{align*}
Let $ p'(t) = z, p(t) = x, p(-t) = y$. Therefore, eq. (1) becomes $$ z = ax + by $$
Let  \begin{align*}
S &= \sum{(z-(ax+by))^2} 
\end{align*}
Differentiating w.r.t a,
\begin{align*}
S &= 2\sum{(z-(ax+by))(-x)} = 0 \\
S &= \sum{(-zx+ax^2+bxy)} = 0\\
\end{align*}
\begin{equation}
\sum{zx}=a\sum{x^2}+b\sum{xy}\\
\end{equation}
Differentiating w.r.t b,
\begin{align*}
S &= 2\sum{(z-(ax+by))(-y)} = 0 \\
S &= \sum{(-zy+axy+by^2)} = 0\\
\end{align*}
\begin{equation}
\sum{zy}=a\sum{xy}+b\sum{y^2}\\
\end{equation}
On solving eq. (11) and eq. (12), we obtain the values of $a$ and $b$.\\
\textbf{ESTIMATING 'A' and 'B':}\\
We have found that  
\begin{align*}
p(t)&= (aA+bB)\exp(-rt) + (aB+bA)\exp(rt)
\end{align*}
Let,
\begin{align*}
p(t)&= y\\
aA+bB &=w1\\
aB+bA &=w2\\
\exp(rt) &= x\\
\end{align*}
Taking log
\begin{align*}
\log(x) &=  rt\\
\log(x^{-1}) &= -rt\\
exp(-rt) &= \frac{1}{x}\\
\end{align*}
Therefore,
\begin{align*}
y &= w1*x + \frac{w2}{x}\\
xy &= w1*x^2 + w2\\
\end{align*}
Let,
\begin{align*}
Y &= xy\\
X &= x^2\\
\end{align*}
Now,
\begin{align*}
Y &=w1*X +w2\\
\end{align*}
\begin{equation}
\sum{Y} =w1\sum{X} +w2*n\\
\end{equation}
\begin{equation}
\sum{X}\sum{Y} =w1\sum{X^2}+w2\sum{X}\\
\end{equation}

On solving eq. (13) and eq. (14) we can obtain values of w1 and w2. Hence,we can also find the values of A and B. We present the algorithm below.
\begin{algorithm}[h]
\caption{Model Fit using Least Square Method}
\begin{algorithmic}[1]
\State $\textit{p(t)} \gets \textit{Input journal influence data}$
\State $\textit{EQU} \gets \textit{Model\_EQ(p(t))}$
\State $\textit{NEW\_EQU} \gets \textit{LSM(EQU)}$
\Procedure{Model\_EQ(p(t))}{}
\State $\textit{p'(t)} \gets \textit{ap(t) + bp(-t)}$
\State $\textit{Discretize the derivative using present and past data}$
\State $\textit{p'(t)} \gets \textit{{p(t+h)-p(t)}/{h}}$
\State $\textit{(p(1)-p(0))}/{1} \gets \textit{ap(-1) + bp(1)}$
\State ${= a[A\exp(-r) + B\exp(r)] + b[A\exp(r)+B\exp(-r)]}$
\State \textit{Return} ${(aA+bB)\exp(-r) + (aB+bA)\exp(r)}$
\EndProcedure
\Procedure{LSM(EQU)}{}
\State \textit{Derive the values of 'a' and 'b' of EQU using Equations}
\State $\sum{zx}=a\sum{x^2}+b\sum{xy}$ \hspace{0.2cm} \textit{and}
\State $\sum{zy}=a\sum{xy}+b\sum{y^2}$
\State \textit{Derive the values of 'A' and 'B' of EQU using Equations}
\State $\sum{Y} =w1\sum{X} +w2*n$ \hspace{0.2cm}
\State $\sum{X}\sum{Y} =w1\sum{X^2}+w2\sum{X}$
\State $return\hspace{0.2cm} \textit{EQU with derived values}$ 
\EndProcedure
\end{algorithmic}
\end{algorithm}
\begin{figure}[h]
 \includegraphics[width=8cm,height=8cm]{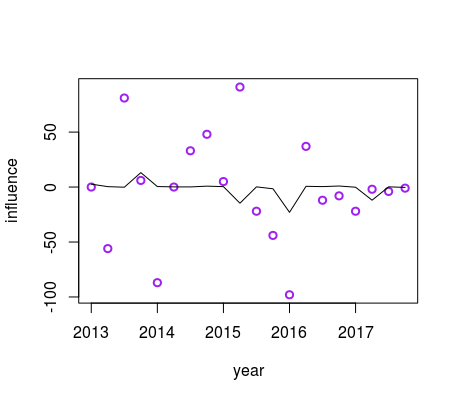}
 \centering \caption{p'(t) VS t  ((Non-linear least square curve): The rate of change in influence over the span of 5 years shows small fluctuations but maintains an overall steady value. This means that the influence in 5 years will not suffer drastically.}
\centering 
\end{figure}
\section{Model Modification to accommodate implicit control variables}
Additionally, we consider implicit control variables which play important roles in the growth of any journal. These variables pose challenges to the modeling set up and without these, the scope is limited to empirical verification at a minor scale only. Next step in modeling data is to carry out modifications to this structure in order to accommodate implicit control parameters such as publisher goodwill value and “start-up initiative” by editors (editorial reputation). We define this initiative as the reputation of editors who steered the journal and offered a strong attraction for quality submissions from scholars across the globe. It is realistic to hypothesize that reputed scholars acting as editors add value and credibility to an emerging journal. This value however is extremely hard to quantify and therefore modeling such phenomenon is novel and imperative to understand the journal’s growth pattern. We propose to present the model and the analytical solution, repeat the exercise of sections 3 and 4 and discuss the implication of the proposed modification.

\par The Time Reversed equation with the additive influence term (Publisher goodwill value) can now be re-written as 
\begin{equation}
p'(t) = ap(t) + bp(-t) + \eta + \theta
\end{equation}
where $\eta$ is an additive term implying goodwill of the publishing house, Elsevier, in our case! $\theta$, OTOH represents Editors' reputation.
\subsection{Additional Considerations}
\begin{itemize}
\item Let us assume $\eta$ to be either linear or exponential. Such considerations are justified since any reputed publisher, in order to remain competitive, would strive to enhance goodwill. Thus, $\eta$ can't possibly be a constant. 
\item We pose the next question pertinent to quantification of goodwill. It is modeled as a function of the percentage of accepted papers over time, a trend that accommodates a fixed number of accepted articles and the selection criteria of additional papers becomes increasingly stringent. It is modeled as
\begin{equation}
\eta(.) = exp(-art) +\alpha (a-b)
\end{equation}
where $art$ is the percentage of articles accepted after the initial threshold of $\alpha$ articles. $\alpha (a-b)$ is the initial threshold, conveniently set to ensure that the influence doesn't hover to the negative.
\item Thus, $\eta(.)$ is a control variable in the formulation and explanation of publisher goodwill. This implies, increasingly the percentage of accepted articles will diminish. Such stringent measures in peer-review bolster publisher goodwill.
\item The formulation being in place, we now integrate $\eta(.)$ with the modified model.
\item Editorial reputation may be any of the three: a constant function, linear or exponential growth. The first one is more likely since the Editors of ASCOM are well established in their fields. Therefore, it is less likely that their phase of influence is still growing at quadratic rate or higher. In fact, we have observed that the influence pattern (citations) is steady. Nonetheless, we have considered all three possibilities and discuss the implications after integrating Editorial reputation, $\theta$ (which is a function of time) in to the model.
\end{itemize}
\subsection{Temporal evolution of publisher goodwill value}
Figures 5(a) and 5(b) throw some useful insights. We hypothesize that the linear graph (fig. 5(b)) is a subset of the non-linear one (fig. 5(a)). Fig 5(b), which is a time-series plot of publisher goodwill value is linear upon fitting the ASCOM data. Fig. 5(a) is an extended time window plot of the same journal which is accomplished by simulating the data available from 5 years, extended to 10 years. The 5-year trend, if we take the time-slice off from fig. 5(a), produces fig. 5(b). This is done to establish the hypothesis that, available data to understand and predict longer time average behavior is insufficient. 
\begin{figure}[t]
\centering
\subfigure[Publisher goodwill v/s time for 10 years]{\includegraphics[width=7cm, height=7cm]{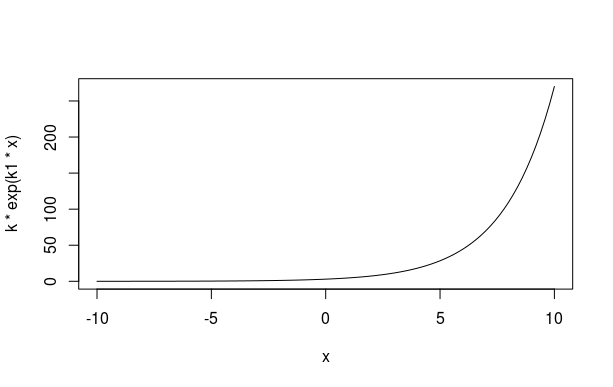}}
\subfigure[Plot of publisher goodwill v/s time in years]{\includegraphics[width=7cm, height=7cm]{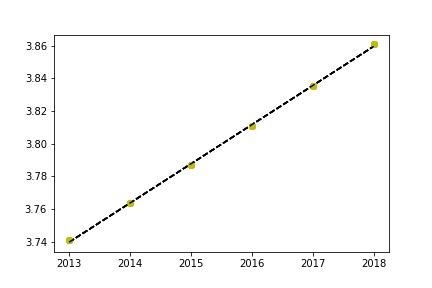}}
\caption{Plot of publishers goodwill VS time. We observe that the publishers good will shows a linear rise in the span of the 5 years between 2013 and 2017. Extrapolated to 10 years, the linear trend becomes non-linear and eventually impact the overall influence of the journal by a margin.}
\label{fig:test}
\end{figure}
\par This synthetic experiment implies that, if the good work in the past continues (good inheritance in terms of positive influence of the implicit control variable, the publisher goodwill, it shall continue to grow in non-linear fashion). The observation is in agreement with the publisher in question, Elsevier, who pursues aggressive and stringent quality practices toward the larger goal of monopoly in the business of publishing. At this point, we may note that, the nonlinear, time dependent trend shall influence the overall journal growth in influence to a greater proportion in comparison with the model we assumed in eq. (16) (which is time-independent). We draw such inferences from the goodwill value as a time series plot by resolving the equation with fitted goodwill value model from time-series data. We show that in the ensuing discussion and the figures below (fig. 6, 7, 8). Let us now consider eq. (16). On adding the publishers goodwill as  a function of time we obtain the equation, \\ 
$p"(t) -(b^2 - a^2)p(t) = (a+b)\theta(t) +k*\exp^{k_1t}$\\ 
On solving the above equation on similar lines outlined in Appendix C, we obtain expressions of journal influence as solutions for the three different cases of $\theta(t)$ being constant, linear and exponential and $\eta(t)$ being the time dependent function instead of a function of accepted articles as discussed earlier.
\begin{enumerate} 
\item
CASE 1 (Fig. 6): Let us assume that $\theta(t) = \theta = constant$\\
$\Longrightarrow p(t) = c_1\exp^{t\sqrt{b^2-a^2}}+c_2\exp^{-t\sqrt{b^2-a^2}} + \frac{\theta}{(a-b)}+\frac{k*exp^{(k_1)*t}}{(k_1)^2-(b^2-a^2)}$
\item
\begin{figure}[h]
\includegraphics[width=7cm,height=7cm]{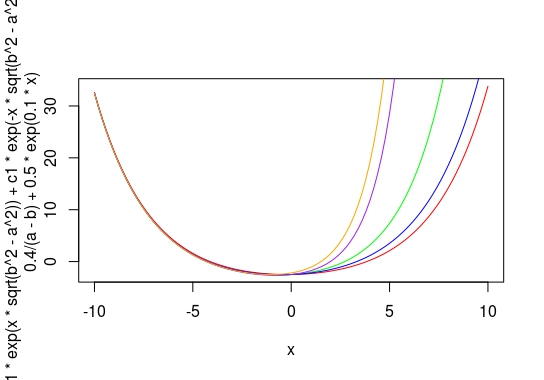}
\centering \caption{Plot of $p(t)$  VS. $t$ when $p(t)$, the influence is the solution to the above equation. We observe that the slope of the graph becomes steeper as $k_1$ increases.}
\centering
\label{fig:rank}
\end{figure}
CASE 2 (Fig. 7): Let us assume that $\theta(t)$ is linear: $\theta(t)$ = At+B \\
$\Longrightarrow p(t) = c_1\exp^{t\sqrt{b^2-a^2}}+c_2\exp^{-t\sqrt{b^2-a^2}} + \frac{At+B}{a-b}+\frac{k*exp^{(k_1)*t}}{(k_1)^2-(b^2-a^2)}$
\begin{figure}[h]
\includegraphics[width=7cm]{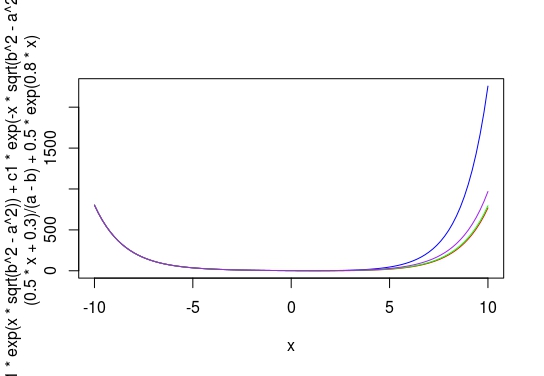}
\centering \caption{Plot of $p(t)$  VS. $t$ when $p(t)$, the influence is the solution to the above equation. We observe that the slope of the graph becomes steeper as $k_1$ increases.
}
\centering
\label{fig:rank}
\end{figure}
\item
CASE 3 (Fig. 8): Let us assume that $\theta(t)$ is exponential:\\$\theta(t) = exp^{At}$ 
$\longrightarrow p(t) = c_1\exp^{t\sqrt{b^2-a^2}}+c_2\exp^{-t\sqrt{b^2-a^2}} + \frac{(a+b)exp^{At}}{A^2-(b^2-a^2)}+\frac{k*exp^{(k_1)*t}}{(k_1)^2-(b^2-a^2)}$
\begin{figure}[h]
\includegraphics[width=7cm]{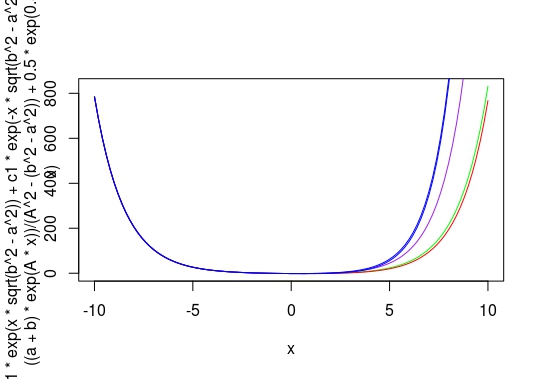}
\centering \caption{Plot of $p(t)$  VS. $t$ when $p(t)$, the influence is the solution to the above equation. We observe that the slope of the graph becomes steeper as $k_1$ increases.
}
\centering
\label{fig:rank}
\end{figure}
\end{enumerate}
These plots (shown in Figure 5, 6, 7 and 8) demonstrate clearly, as publisher goodwill value is modeled as a time dependent evolution, the influence of the journal grows at a faster pace in the longer run. Therefore, it complements our observation that, publisher goodwill value has a small role to play in the growth of journal influence in short time span but evolves gradually as time elapses.
\subsection{Temporal evolution of Editorial reputation}
We observe the celebrity effect here (\citet{articleFei}). Editors are well established scholars and by the time they assumed editorial responsibility, they are in the "cool off state" implying the surge in reputation they experienced when they were rising stars stabilized. Therefore, steep gradient shall no longer be expected. This is what we observe in Fig 9 where the editorial influence between 2004 and 2014 is plotted. Please note ASCOM was founded in 2013. The influence trend of all the editors during that time (2010-14) is approximately constant.
\begin{figure}[h]
 \includegraphics[width=7cm,height=7cm]{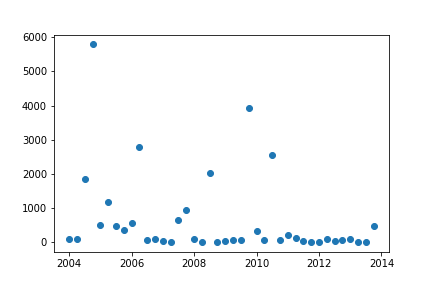}
 \centering \caption{The above plot represents editors influence VS time. We see that the editorial influence is almost constant with time. This is possible because the editors are already well established. Hence, the influence is steady with little fluctuations.}
\centering 
\end{figure}
\par Next section will deliberate on the contributions of these variables, in particular and model modification, in general on the rate of change in influence observed in ASCOM. The role of control variables are evident in the visualization we present below.

\section{DISCUSSION}
We develop a model to study its effect on astronomy and computer science domains and analyze parameters that have contributed in building the reputation of ASCOM. In this specific case study of journal influence, the spread is clearly dependent on present as well as history dependent functions. This strengthens the motivation of using DDE model for the study. The model explains the growth pattern of the journal well by capturing the intrinsic attributes and historical data. The time reversed model works as a mirror and helps carry over the good deeds of the past (quality of articles in niche areas and open problems solved by interdisciplinary efforts reflected in citation history). Our model takes care of the “hereditary effects” and since the phenomenon of observing a journal in an emerging and interdisciplinary area is modeled as a function of spatial variables renders the system infinite degrees of freedom. Thus, the proposed model is robust and provides better control over the system.
\par However, the data is limited since the journal is in publication for just over five years. Therefore the influence of historical data does not translate to overwhelming quantitative evidence in the way we liked it to. Nonetheless, if we extrapolate the interval by extending the time window of consideration (since the historical data is assumed to influence the present one), we observe profound effects (Please see the discussion on temporal evolution of publisher goodwill value where the observed linear growth in goodwill is really a 5-year snapshot subset of the longer window; (please see figs. 5(a) and (b) and the discussion in section 6.2). Additionally, we considered implicit control variables such as Editorial reputation and Publisher goodwill which play important roles in the growth of any journal. These variables pose challenges to the modeling set up and without those, the scope is limited to empirical verification at minor scale. We observe,
\begin{itemize}
\item The graphs for the equations for $p(t)$ vs time where $Y$ axis represents $p(t)$ and $X$ axis is time has a parabolic shape. This shape is due to the presence of the symmetric history function $p(-t)$.
\item From the results of this study, it can be established that the journal citations vary in a non-linear fashion. Initially, the citation score is usually less as the journal will have less popularity. This can be seen when we do a comparison of the citations of the editor v/s time and the influence of the journal v/s time.

Fig 9. shows that the rate of change in influence is more or less constant over time. There is an initial irregularity as the initial change in influence is directly related to the current influence. But with time, the graph smooths out because, the other parameters such as citations and readership of the journal also begin to affect the rate of change of influence. This is a testimony of consistent and largely positive rate of change in influence. This  hypothesis is verified by the graphs discussed below.
First, we plot the the editorial influence with time without considering the publishers goodwill (as time series data) and editorial influence. We see that the curve is a simple parabola. This confirms our assumption that the influence has a global minima and it stays upward elsewhere.
\item After adding publishers Goodwill which we assumed to be an exponential distribution, the plot shows a small change in shape. Since publisher's goodwill is a constant followed by a -ve exponential function, it can be conveniently added to $a p(t)+b p(-t)$, as those are also exponential. This shows that the publishers goodwill plays a small but non-negligible role in governing the journals growth in influence. However, in line with our argument presented in section 6, the small change will translate to larger increment if the time window is expanded i.e. the goodwill will have a larger contribution with the elapse of time.
On plotting the curve for influence VS time after considering different Editorial influence functions (i.e. $\theta$(t)) along with the publishers goodwill, we obtain the following graphs (fig. 10.(a), (b) and (c)).
\item
\begin{figure}[h]
\centering
\subfigure[$p(t)$ (i.e journal influence) VS $t$ is positive, including the initial influence. This is because of the "start-up boost" provided by the control variables.]{\includegraphics[width=6cm, height=6cm]{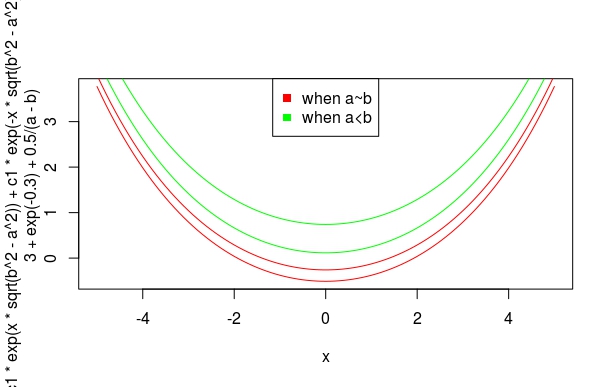}} \hspace{1cm}
\subfigure[p(t) VS t when$\theta(t)$ (i.e editorial influence) is linear. Initial influence is shifted further up.]{\includegraphics[width=7cm, height=6cm]{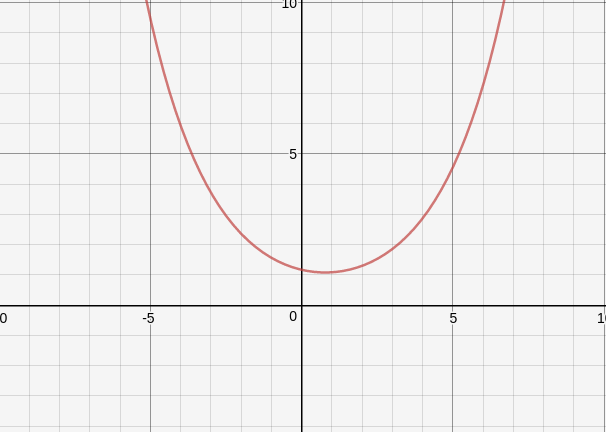}}
\subfigure[p(t) VS t when $\theta(t)$ (i.e editorial influence) is exponential]{\includegraphics[width=6cm, height=6cm]{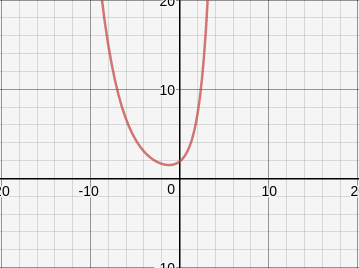}}
\caption{Journal Influence $p(t)$ v/s $t$ for different variations in $\theta$}
\label{fig:test}
\end{figure}
Note, $\eta/(a-b) >= \alpha$ when $c_1 = c_2$. If this condition is not satisfied the values of the influence becomes negative for initial values of time which is not possible. The model ensures that.
When the editorial influence is exponential it gets added to the positive part of the delay differential equation.

\item We simulated/computed goodwill value, $\eta$ for every year. The value changes, as expected and therefore, $\eta$ can now be interpreted as time series data, fitted as a function of time and used in the modified equation as a time-dependent function. With this modification, we resolve the DDE equation and noted changes in the trend. The gradient (rate of change in influence) is greater as compared to the earlier model where $\eta$ is considered as a function of percentage of accepted articles!
\item The magnitudes of the constants $c_1$ and $c_2$ determine if the recent or the historic part of the equation dominates the curve.
\item \textbf{Effect of implicit control variables}: Editorial reputation and publisher goodwill value are indeed positive factors. When the figures are compared, (figs. 2, 6, 7 and 10), we observe a non-zero start-up boost in figs. 6, 7, 8 and 10. On the contrary, when we investigated the original DDE (fig. 2) without these implicit variables, influence evolution begins from 0. This non-negative boost (graphs of figs. 8, 9 and 10 start from some positive value above 0) is due to influence of the control variables, which are otherwise hard to quantify and model. Therefore, our arguments in favor of including these variables in the influence model stand vindicated.
\end{itemize}
\section{CONCLUSION} model suitably explains the growth pattern of the journal by capturing the intrinsic attributes and historical data. We observe effects of celebrity authorship in the role of editors contribute to the growth in influence of ASCOM s well as the goodwill established publishing house brings. These effects, dynamic in nature, haven't been studied before. The contribution of the manuscript is therefore two-fold. Firstly, a novel model of DDE is exploited to study the influence of a journal in an emerging area. Secondly, qualitative and dynamic control variables (Editorial reputation and publisher goodwill value), hitherto unexploited, for the simple reason of complexity have been quantified and integrated in to the model. The time reversed model works as a mirror and helps to carry over the achievements of the past (quality of articles in niche areas and open problems solved by interdisciplinary efforts). As a final note, it might be useful to remind that our model
takes care of the “hereditary effects” by exploiting the function, $ p(-t)$. The phenomenon of observing a journal in an emerging and interdisciplinary area is modeled as a function of spatial variables renders the system infinite degrees of freedom. Therefore, the proposed model is robust and it provides better control over the system.
Note however, that the data is limited given the low age of the journal. Therefore the influence of historical data does not translate into overwhelming quantitative evidence. The model also holds promise because of its control structure and ability to accommodate implicit control variables such as Editorial reputation and publisher goodwill value have been found to generate significant implications, overall. We establish a fact, no less remarkable, that the implicit control variables act as incentives to the new journal in an emerging area. This is a much needed boost that the journal enjoys, in the absence of which, it may have to struggle much harder to attract readership and scholarship! We conclude by stating that unlike most of the scholarly work in scientometric literature, which are post-facto statistical studies, our work is focused on investigating the background responsible for certain trends observed in a journal in niche area. This is where the manuscript is markedly different!
\section{Acknowledgements}

We would like to thank the Science and Engineering Research Board (SERB)-Department of Science and Technology (DST), Government of of India for supporting our research. The project reference number is: SERB-EMR/2016/005687.

\bibliographystyle{apacite}  	
\bibliography{myRef} 		

\section*{Appendix}

\appendix
\setcounter{table}{0}		
\setcounter{figure}{0}		
\renewcommand{\thetable}{\Alph{section}.\arabic{table}}		
\renewcommand{\thefigure}{\Alph{section}.\arabic{figure}} 	


\section{First Appendix}
  \label{apdx:1st}
\subsection{\emph{Singular Value Decomposition}}
 
Singular Value Decomposition is the factorization of a real or complex matrix. Large scale of Scientometric data is mined using suitable web scraping techniques and is modeled as a matrix in which the rows represent the articles in a journal published over the years, and the columns represent various Scientometrics or indicators proposed by experts of evaluation agencies (\citet{Kalman96}). The original data matrix, say \textbf{A} of dimension \textit{m}x\textit{n} and rank \textit{k} is factorized into three unique matrices \textbf{U}, \textbf{V} and \textbf{W\textsuperscript{H}}. 
\begin{itemize}
\item \textbf{U} - Matrix of Left Singular Vectors of dimension \textit{m}x\textit{r}
\end{itemize}
\begin{itemize}
\item \textbf{V} - Diagonal matrix of dimension \textit{r}x\textit{r} containing singular values in decreasing order along the diagonal
\end{itemize}
\begin{itemize}
\item \textbf{W\textsuperscript{H}} - Matrix of Right Singular Vectors of dimension \textit{n}x\textit{r}. The Hermitian, or the conjugate transpose of \textbf{W} is taken, changing its dimension to \textit{r}x\textit{n} and hence the original dimension of the matrix is maintained after the matrix multiplication.
In this case of Scientometrics, since the data is represented as a real matrix, Hermitian transpose is simply the transpose of \textbf{W}.
\end{itemize}

\textit{r} is a very small number numerically representing the approximate rank of the matrix or the number of "concepts" in the data matrix \textbf{A}. \textit{Concepts} refer to latent dimensions or latent factors showing the association between the singular values and individual components (\citet{Kalman96}). The choice of \textit{r} plays a vital role in deciding the accuracy and computation time of the decomposition. If \textit{r} is equal to \textit{k}, then the SVD is said to be a Full Rank Decomposition of \textbf{A}. Truncated SVD or Reduced Rank Approximation of \textbf{A} is obtained by setting all but the first \textit{r} largest singular values equal to zero and using the first \textit{r} columns of \textbf{U} and \textbf{W} (\cite{Apache2017}). \par
Therefore, choosing a higher value of \textit{r} closer to \textit{k} would give a more accurate approximation whereas a lower value would save a lot of computation time and increase efficiency.
\subsection{\emph{Regularization Norms}}

In the case of Big Data, parsimony is central to variable and feature selection, which makes the data model more intelligible and less expensive in terms of processing.

\textit{l$_p$}-norm of a matrix or vector \textbf{x}, represented as \(||\)\textbf{x}$_p$\(||\) is defined as, \(||\)\textbf{x}$_p$\(||\) = $\sqrt[p]{\Sigma_i{|x|_i}^p}$
i.e the p\textsuperscript{th} root of summation of all the elements raised to the power p.
Hence, by definition,  \textit{l$_1$} norm = \(||\)\textbf{x}\(||\)$_1$ = $\Sigma_i{|x|_i}$

Sparse approximation, inducing structural sparsity as well as regularization is achieved by a number of norms, the most common ones being \textit{l$_1$} norm and the mixed group \textit{l$_1$-l$_q$} norm. 
The relative structure and position of the variable in the input vector, and hence the inter-relationship between the variables is inconsequential as a variable is chosen individually in \textit{l$_1$} regularization. 
Prior knowledge aids in improving the efficacy of estimation through these techniques. \par
The \textit{l$_1$} norm concurs to only the cardinality constraint and is unaware to any other information available about the patterns of non-zero coefficients.(\citet{Bach2012})

\subsection{\emph{Sparsity via the \textit{l$_1$} norm}}

Most variable or feature selection problems are presented as combinatorial optimization problems. Such problems focus on selecting the optimal solution through a discrete, finite set of feasible solutions.  Additionally, \textit{l$_1$} norm turns these problems to convex problems after dropping certain constraints from the overall optimization problem. This is known as convex relaxation. Convex problems classify as the class of problems in which the constraints are convex functions and the objective function is convex if minimizing, or concave if maximizing. 
\par \textit{l$_1$} regularization for sparsity through supervised learning involves predicting a vector \textbf{y} from a set of usually reduced values/observations consisting a vector in the original data matrix \textbf{x}. This mapping function is often known as the hypothesis \textbf{h : x$\to $y}. To achieve this, we assume there exists a joint probability distribution P(x,y) over \textbf{x} and \textbf{y} which helps us model anomalies like noise in the predictions. 

\par In addition to this, another function known as a loss function L(y',y) is required to measure the difference in the prediction y'=h(x) from the true result y. Consider the resulting vectors consisting of the predicted value and the true value to be \textbf{y'} and \textbf{y} respectively. A characteristic called \textit{Risk}, R(h) associated with loss function, and hence in turn with the hypothesis-h(x) is defined as the expectation of the loss function.
$$R(h) = \textbf{E}[L(y',y)] = \int L(y',y)\,dP(x,y)$$ 
Thus, the hypothesis chosen for mapping should be such that the risk, R(h) is minimum. This refers to as risk minimization.
However, in usual cases, the joint probability distribution of the problem in hand, P(x,y) is not known. So, an approximation called \textit{empirical risk} is computed by taking the average of the loss function of all the observations. Empirical Risk is given by : $$R_{emp}(h)=\frac{1}{n} \sum_{i=1}^{n}L(\textbf{y'}_i,\textbf{y}_i)$$
The empirical risk minimization principle states that the hypothesis(h') selected must be such it that reduces the empirical risk $R_{emp}(h)$:
$$h'=\min_{h}R_{emp}(h)$$ 

While mapping observations x in \textit{n} dimensional vector \textbf{x} to outputs y in vector \textbf{y}, we consider \textit{p} pairs of data points - $(\textbf{x}_i,\textbf{y}_i) \in \mathbb{R}^n \times \textbf{y}$ where i = 1,2...p. 
Thus the optimization problem for the data matrix in Scientometrics takes the form: $$\min_{\textbf{w}\in\ \mathbb{R}^n} \frac{1}{p}\sum_{i=1}^{p}L(\textbf{y'}_i,\textbf{w}^T\textbf{x}_i) + \lambda\Omega(\textbf{w})$$ 
L is a loss function which can either be square loss for least squares regression, $L(y',y)=\frac{1}{2}(y'-y)^2$, or a logistic loss function. Now, the problem thus takes the form: $$\min_{\textbf{w}\in \mathbb{R}^n}||\textbf{y'}-\textbf{A}\textbf{w}||^2$$
Since the variables in the vector space/groups can overlap, it is ideal to choose $\Omega(\textbf{w})$ to be a group norm for better predictive performance and structure. The \textit{m} rows of data matrix \textbf{A} are treated as vectors or groups(\textit{g}) of these variables, forming a partition equal to the vector dimension, [1:\textit{n}]. If \textbf{\textit{G}} is the set of all these groups and \textit{d$_g$} is a scalar weight indexed by each group \textit{g}, the norm is said be a \textit{l$_1$}-\textit{l$-q$} norm where $q\in[2,\infty)$ (\citet{Bach2012}), $\Omega(\textbf{w}) = \sum_{g\in\textbf{\textit{G}}} \textit{d$_g$}||\textbf{w}_g||_q$\\  
The choice of the indexed weight \textit{d$_g$} is critical because it is responsible for the discrepancies of sizes between the groups. It must also compensate for the possible penalization of parameters which can increase due to high-dimensional scaling. The factors that affect the selection are the choice of q in the group norm and the consistency that is expected of the result. In addition to this, accuracy and efficiency can be enhanced by weighing each coefficient in a group rather than weighing the entire group as a whole.
The initial sparse data matrix is first manipulated using the \textit{l$_1$}-norm (\citet{Bach2012}).
\subsection{Methodology}
An estimate of a journal's scholastic indices is necessary to judge its effective impact. The nuances of scientometric factors such as Total Citation Count and Self-citation Count come into play when deciding the impact of a journal. However, these factors unless considered in ideal circumstances don't by themselves become a good indicator to represent the importance of a journal. Many anomalies arise when considering these indices directly which may misrepresent or falsify a journal's true influence. The necessity to use these indices in context with a ranking algorithm is imperative to better utilize these indices. The resulting transformation of $l_1$-norms gives rise to a row matrix which is of the length equal to the number of features of the pristine Scientometric data. This row matrix effectively represents the entire dataset at any given iteration. The application of the Singular Value Decomposition operation on this row matrix is key in determining the necessary norm values to remove through a recursive approach. 

The $singval$ array contains the Normalized Singular Values of all the individual $l_1$-norm transformed columns. These values act as scores while addressing the impact of any given journal. In the context of Singular Values the one with the lowest $singval$ score is the most influential journal.

\begin{algorithm}
\caption{Recursive $l_1$-norm SVD}
\begin{algorithmic}[1]
\State $\textit{A} \gets \textit{Input Transposed Feature Matrix A}$
\Procedure{Lasso}{}
\State $\textit{row\_matrix} \gets \text{Coefficents of }\textit{Lasso Regression}$
\State $return\hspace{0.3cm} \textit{row\_matrix}$
\EndProcedure
\Procedure{SVD}{}
\State $\textit{U,$\Sigma,$V} \gets \text{Matrices of }\textit{SVD}$
\State $return\hspace{0.2cm} \Sigma$ 
\EndProcedure
\Procedure{Normalize}{}
\State $\textit{Norm\_Data} \gets \text{Normalized using }\textit{$l_1$-norms}$
\State $return\hspace{0.2cm} \textit{Norm\_Data}$ 
\EndProcedure
\Procedure{Recursive}{}
\State $\textit{L1\_row} \gets LASSO(A)$
\State $\textit{singval []} \gets SVD(L1\_row)$
\State $\textit{Row\_Norm} \gets Normalize(L1\_row)$
\State $\textit{Col\_Norm} \gets Normalize(\text{All columns of A})$
\State $\textit{Col\_i} \gets \text{Closest Col\_Norm Value to Row\_Norm} $
\State $\text{\textbf{Delete} Col\_i from A}$ 
\State $goto\hspace{0.2cm} \textit{RECURSIVE}$ 
\EndProcedure
\end{algorithmic}
\end{algorithm}
\newpage
\section{Lemma : The equation  $p'(t) =a p(-t) +b p(t)$, 
where $p(0)=K$ and $p'(0)=(a+b)K$ has a unique solution in $[-d,d]$.} 
Proof: Let us consider two solutions to this equation, 
$y_1(t)$, $ y_2(t)$. Then,
\begin{align*}
g(t) &= y_1(t) - y_2(t)\\
y_1'(t) &= ay_1(-t) + by_1(t)\\
y_2'(t) &= ay_2(-t) + by_2(t)\\
\Longrightarrow (y_1-y_2)'(t) &= a(y_1-y_2)(-t) + b(y_1-y_2)(t)\\
\Longrightarrow g'(t) &= ag'(-t) +bg(t)\\
g'(0) &= ag(0) + bg(0) \neq 0\\
g(0) &= y_1(0) - y_2(0)\\
&= k - k = 0\\
\Longrightarrow g'(t) &= ag(-t) + bg(t) \\
g(0) = 0\\
\Longrightarrow g(t) = 0  \forall  t \in [-d,d]\\
\Longrightarrow y_1(t) = y_2(t) \forall t \in [-d,d]
\end{align*}
Therefore, the DDE has a unique solution, which implies that if there exist a oscillatory solution, there can not be an exponential or linear family of solutions depending on the parameters.
\section{DDE solution}
Let us consider a non-linear homogeneous DDE:
$ p"(t) -(b^2 - a^2)p(t) = (a+b)\theta(t)$
The solution for this equation depends on the value of $\eta$.
Auxiliary equation is: 
\begin{align*}
AE &= (D^2 - (b^2-a^2))p(t) = 0 \\
AE &= (m^2 - (b^2-a^2))p(t) = 0 \\
m &= (b^2-a^2)^(1/2)\\
c_y = c_1\exp^{t\sqrt{b^2-a^2}}+c_2\exp^{-t\sqrt{b^2-a^2}}\\
\mathrm{Since}~\sqrt{b^2-a^2}=r, c_y &= c_1\exp^{rt}+c_2\exp^{-rt}\\
\mathrm{The~PI~(particular~integral)~is~calculated~as~follows }\\
\mathrm{CASE 1: Let~us~assume~that~}\theta(t) = \theta = constant\\
y_p &= \frac{(a+b)\theta}{D^2-(b^2-a^2)}\\
y_p &= \frac{exp^{0}(a+b)\theta}{D^2-(b^2-a^2)}\\
y_p &= \frac{(a+b)\theta}{0^-(b^2-a^2)}\\
y_p &= \frac{(a+b)\theta}{-(b-a)(b+a)}\\
y_p &= \frac{\theta}{(a-b)}\\
\Longrightarrow p(t) &= c_1\exp^{t\sqrt{b^2-a^2}}+c_2\exp^{-t\sqrt{b^2-a^2}} + \frac{\theta}{(a-b)}\\
\mathrm{CASE 2: Let~us~assume~that~\eta(t)~is~linear:}~\eta(t) = At+B \\
y_p &= \frac{(a+b)( At+B)}{D^2-(b^2-a^2)}\\
y_p &= \frac{(a+b)( At+B)}{(b^2-a^2)(\frac{D^2}{(b^2-a^2)}-1)}\\
y_p &= \frac{(a+b)}{a^2-b^2}*(1-\frac{D^2}{(b^2-a^2)})^{-1}*(At+B)\\
y_p &= \frac{(a+b)}{a^2-b^2}*(1+\frac{D^2}{(b^2-a^2)}+\frac{D^4}{(b^2-a^2)^2})*(At+B)\\
y_p &= \frac{ At+B}{a-b}\\
\Longrightarrow p(t) &= c_1\exp^{t\sqrt{b^2-a^2}}+c_2\exp^{-t\sqrt{b^2-a^2}} + \frac{At+B}{a-b}\\
\mathrm{CASE 3: Let~us~assume~that~\theta(t)~is~expenetial:}~\theta(t) = exp^{At} \\
y_p &= \frac{(a+b)exp^{At}}{D^2-(b^2-a^2)}\\
y_p &= \frac{(a+b)exp^{At}}{D^2-(b^2-a^2)}\\
y_p &= \frac{(a+b)exp^{At}}{A^2-(b^2-a^2)}\\
\Longrightarrow p(t) &= c_1\exp^{t\sqrt{b^2-a^2}}+c_2\exp^{-t\sqrt{b^2-a^2}} + \frac{(a+b)exp^{At}}{A^2-(b^2-a^2)}\\
\end{align*}
\end{document}